\documentclass[12pt]{article}
\usepackage[colorlinks=false,pdfborder={0 0 0}]{hyperref}
\usepackage{mciteplus}
\usepackage{graphicx}
\usepackage{color}
\usepackage{relsize}
\usepackage{bm}
\usepackage{xspace}
\usepackage{ifthen}
\newcommand\pubnumber{DPF2015-143}
\newcommand\pubdate{October 31, 2015}

\newboolean{articletitles}
\setboolean{articletitles}{true}
\newboolean{inbibliography}
\setboolean{inbibliography}{false}
\def\Title#1{\begin{center} {\Large #1 } \end{center}}
\def\Author#1{\begin{center}{ \sc #1} \end{center}}
\def\Address#1{\begin{center}{ \it #1} \end{center}}

\newcommand\pubblock{\rightline{\begin{tabular}{l} \pubnumber\\
         \pubdate  \end{tabular}}}
\newenvironment{Abstract}{\begin{quotation}  }{\end{quotation}}
\newenvironment{Presented}{\begin{quotation} \begin{center} 
             PRESENTED AT\end{center}\bigskip 
      \begin{center}\begin{large}}{\end{large}\end{center} \end{quotation}}

\def\beq{\begin{equation}}
\def\eeq#1{\label{#1}\end{equation}}
\def\eeqn{\end{equation}}

\def\beqa{\begin{eqnarray}}
\def\eeqa#1{\label{#1}\end{eqnarray}}
\def\eeqan{\end{eqnarray}}

\let\bar=\overbar

\def\Dslash{\not{\hbox{\kern-4pt $D$}}}
\def\dslash{\not{\hbox{\kern-2pt $\del$}}}

\def\msb{{\bar{\ssstyle M \kern -1pt S}}}

\newcommand{\executeiffilenewer}[3]{%
  \ifnum\pdfstrcmp{\pdffilemoddate{#1}}%
  {\pdffilemoddate{#2}}>0%
  {\immediate\write18{#3}}\fi%
}

\newcommand{\tev}{\ensuremath{\mathrm{\,Te\kern -0.1em V}}\xspace}
\newcommand{\gev}{\ensuremath{\mathrm{\,Ge\kern -0.1em V}}\xspace}
\newcommand{\mev}{\ensuremath{\mathrm{\,Me\kern -0.1em V}}\xspace}
\newcommand{\ifb}{\ensuremath{\mbox{\,fb}^{-1}}\xspace}
\newcommand{\pt}{\ensuremath{p_\mathrm{T}}\xspace}
\newcommand{\ip}{\ensuremath{\mathrm{IP}}\xspace}
\newcommand{\bdtl}{\ensuremath{\mbox{BDT}(bc|udsg)}\xspace}
\newcommand{\bdth}{\ensuremath{\mbox{BDT}(b|c)}\xspace}
\newcommand{\sv}{\ensuremath{\mathrm{SV}}\xspace}
\newcommand{\jet}{\ensuremath{\mathrm{j}}\xspace}

\newcommand{\format}{1}
\newcommand{\ifdraft}[2]{\ifthenelse{\equal{\format}{0}}{#1}{#2}}
\newcommand{\ifarxiv}[2]{\ifthenelse{\equal{\format}{1}}{#1}{#2}}

\ifdraft{\usepackage{lineno}\linenumbers}{}

\begin{document}
\begin{titlepage}
\pubblock

\vfill
\Title{Forward $W+c,b$-jet and Top Measurements with LHCb}
\vfill
\Author{Philip Ilten}
\Address{Massachusetts Institute of Technology\\ 77 Massachusetts
  Avenue, Cambridge, MA 02139, USA}
\vfill
\begin{Abstract}
Inclusive $c$ and $b$-jet tagging algorithms have been developed to
utilize the excellent secondary vertex reconstruction and resolution
capabilities of the LHCb detector. The validation and performance of
these tagging algorithms are reported using the full run 1 LHCb
dataset at $7$ and $8\tev$. Jet-tagging has been applied to $\mu+$jet
final states to measure both the $W+c,b$-jet charge asymmetries and
the ratios of $W+c,b$-jet to $W+$jet and $W^\pm+$jet to $Z+$jet
production. The forward top production cross-section is also measured
using the $\mu+b$-jet final. All results are found to be consistent
with standard model predictions.
\end{Abstract}
\vfill
\begin{Presented}
DPF 2015\\
The Meeting of the American Physical Society\\
Division of Particles and Fields\\
Ann Arbor, Michigan, August 4--8, 2015\\
\end{Presented}
\vfill
\end{titlepage}
\def\thefootnote{\fnsymbol{footnote}}
\setcounter{footnote}{0}

\section{Introduction}

Inclusive identification of jets originating from $c$ and $b$-quarks
is a critical experimental technique needed for both standard model
(SM) measurements such as the top cross-section, and beyond the
standard model (BSM) searches, \textit{e.g.} axigluon searches via
charge asymmetries in di-$b$-jet production. LHCb is a forward arm
spectrometer~\cite{Alves:2008zz} located on the large hadron collider
(LHC), with a pseudo-rapidity range of $2 < \eta < 5$ and initially
designed to measure properties of $b$-hadron decays. Heavy-flavor jets
typically contain secondary vertices from $c$ and $b$-hadron
decays. With excellent secondary vertex reconstruction, LHCb is an
ideal environment for $c,b$-jet tagging, while its forward coverage
provides complementary results to the general purpose detectors on the
LHC.

In these proceedings the LHCb $c,b$-jet tagging and its application to
physics measurements in run 1 LHCb data is reported. Two main datasets
recorded at different $pp$ collision energies are used here, a
$1~\ifb$ dataset at $\sqrt{s} = 7\tev$ recorded in $2011$ and a
$2\ifb$ dataset at $\sqrt{s} = 8\tev$ recorded in $2012$. In
Section~\ref{sec:tag} the $c,b$-jet tagging algorithm, validation, and
performance of~\cite{LHCb-PAPER-2015-016} is outlined. Ratios of
$W+c,b$-jet and $Z+$jet production using $\mu+$jet final states
from~\cite{LHCb-PAPER-2015-021} are presented in
Section~\ref{sec:wcb}. Finally, a subsample of $\mu+b$-jet final state
events with a tightened kinematic region is used to measure forward
top-quark production cross-sections~\cite{LHCb-PAPER-2015-022} in
Section~\ref{sec:top}.

\section{Heavy-Flavor Jet Tagging}\label{sec:tag}

In~\cite{LHCb-PAPER-2015-016}, heavy-flavor jets are tagged using
$n$-body secondary vertices (SV) built from the tracks of charged
particles. Two-body SVs are created from displaced tracks in the event
with transverse momentum, \pt, $> 0.5\gev$, and must pass basic
quality requirements. All SVs with shared tracks are combined to
produce $n$-body SVs, such that tracks are unique to a single
SV. Additional loose requirements, consistent with $c$ and $b$-hadron
decays, are applied to these SVs. A jet is heavy-flavor tagged if the
$\Delta R \equiv \sqrt{\Delta \phi^2 + \Delta \eta^2}$ between the
flight-direction for a SV and the jet momentum is less then the jet
radius parameter, $R$. For these studies jets are built from particle
flow input~\cite{LHCb-PAPER-2013-058} with $R = 0.5$ using the
anti-$k_\mathrm{T}$ algorithm~\cite{Cacciari:2008gp}.

For each SV-tag the responses of two boosted decision trees (BDTs) are
calculated; \bdtl discriminates light-jets from $c,b$-jets and \bdth
separates $c$-jets from $b$-jets. Ten variables are used as input to
the BDTs: SV mass, corrected mass\footnote{Corrected mass is defined
  as $\sqrt{m^2 + \pt^2} + \pt$ where \pt is the missing momentum
  transverse to the SV flight-direction and $m$ is the mass of the
  SV.}, transverse flight distance, $\pt(\sv)/\pt(\jet)$, $\Delta R$
between the jet momentum and \sv flight-direction, number of tracks in
the SV, number of SV tracks with $\Delta R < 0.5$ to the jet, net
charge of the SV, flight-distance $\chi^2$, and summed $\ip_{\chi^2}$
of the tracks in the SV. Example two-dimensional distributions from
simulation of the \bdtl and \bdth responses for light, $c$, and
$b$-jets are given in Figure~\ref{fig:tag:templates}. Light jets
cluster at the origin, $c$-jets in the lower right, and $b$-jets in
the upper right.

\begin{figure}
  \begin{center}
    \includegraphics[height=4.5cm]{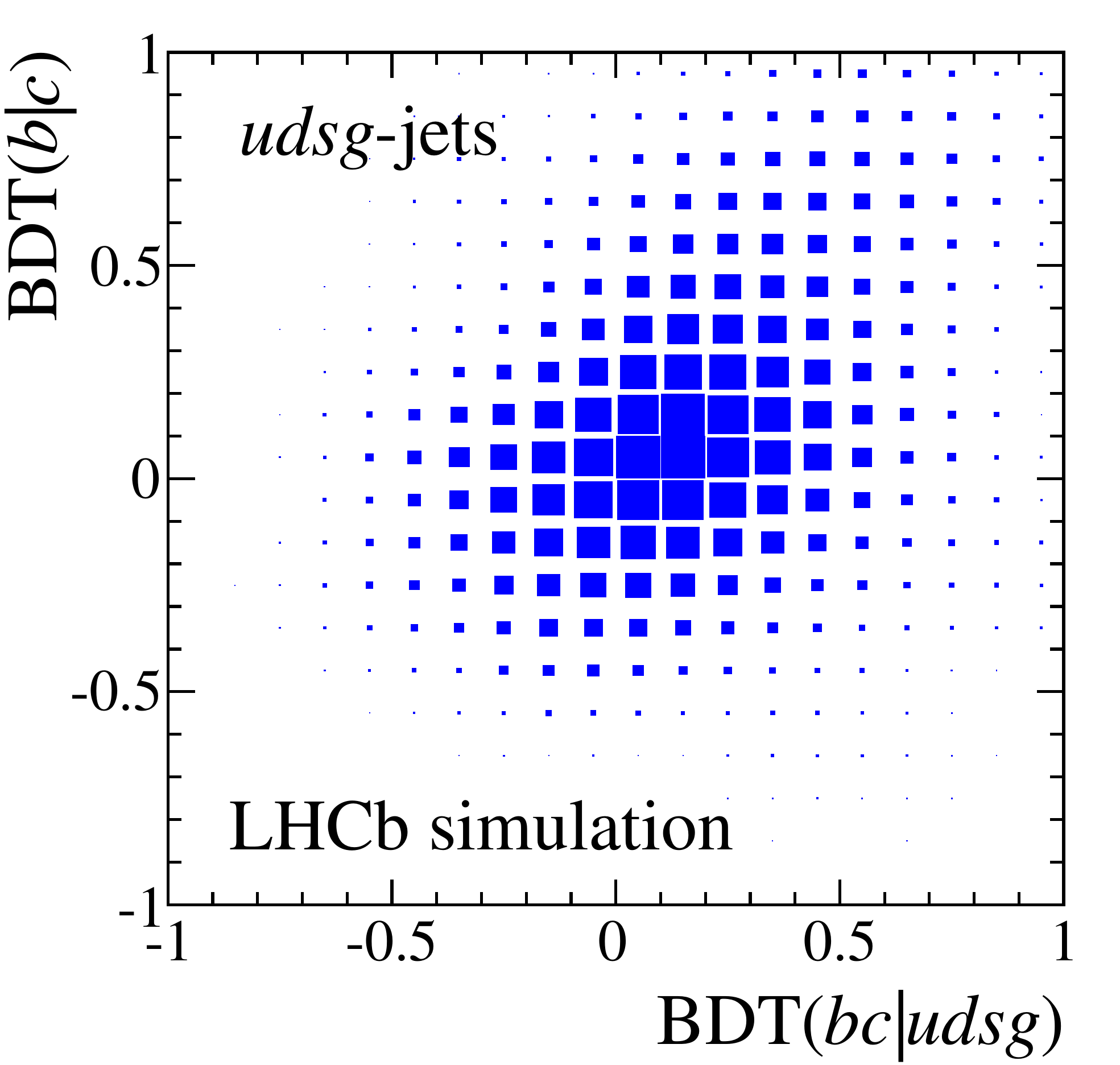}
    \includegraphics[height=4.5cm]{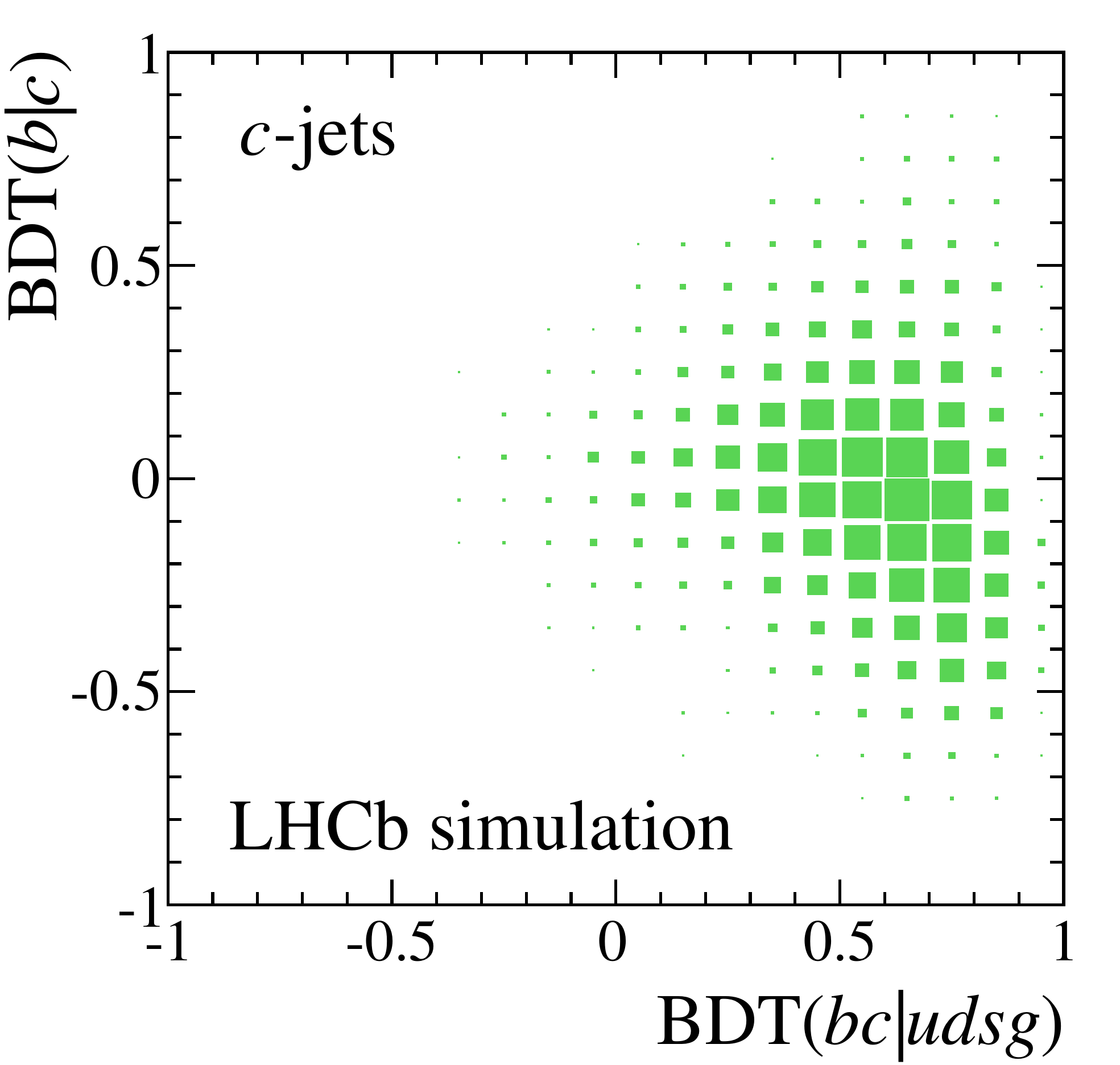}
    \includegraphics[height=4.5cm]{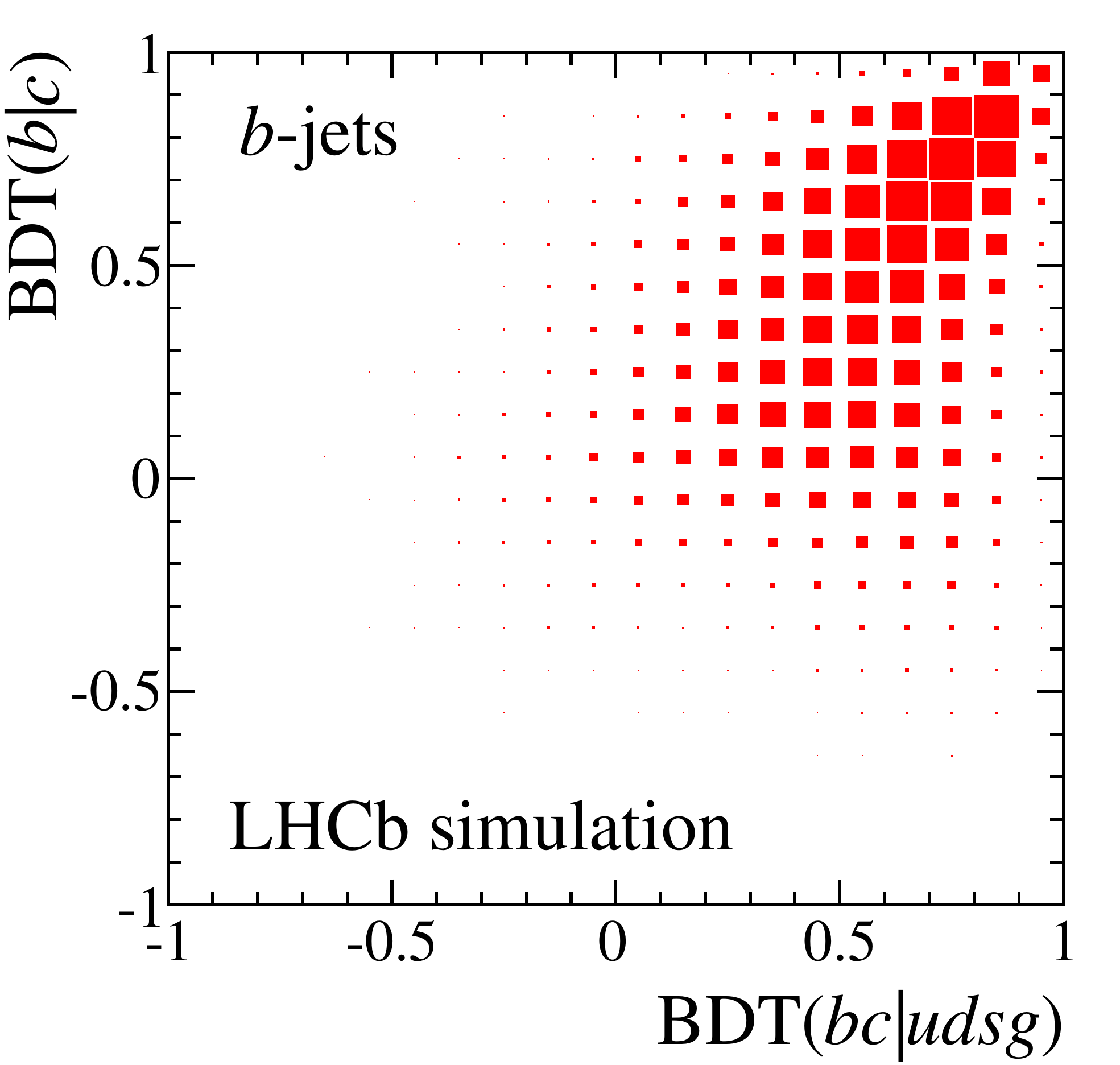}
  \end{center}
  \caption{Two-dimensional distributions from LHCb simulation as a
    function of \bdtl and \bdth responses for (left) light, (middle)
    $c$, and (right) $b$-jets.\label{fig:tag:templates}}
\end{figure}

Four data samples are used for efficiency determination with the
tag-and-probe method. Three are di-jet samples containing a tag-jet
with a fully reconstructed $b$-hadron, a fully reconstructed
$c$-hadron, or a displaced muon. The probe-jets of the $b$-hadron
sample are $b$-enriched, while the probe-jets of the $c$-hadron and
displaced muon samples are both $c$ and $b$-enriched. The fourth
sample requires an isolated high-\pt tag-muon and a probe-jet, which
is light-jet enhanced. The jet-tagging efficiency in each sample is
the number of tagged probe-jets over the total number of probe-jets
for a given jet type (light, $c$, or $b$).

The probe-jet flavor composition prior to SV-tagging is determined by
fitting the $\log(\ip_{\chi^2})$ distributions for the hardest-\pt
track or hardest \pt-muon of the jet. After SV-tagging, the probe-jet
flavor is determined by a two-dimensional fit of the \bdtl and \bdth
response distribution. Projections of this fit onto the \bdth and
\bdtl axes for the $c$-hadron sample are shown in
Figure~\ref{fig:tag:fits}.

\begin{figure}[t!]
  \begin{center}
    \includegraphics[height=4.5cm]{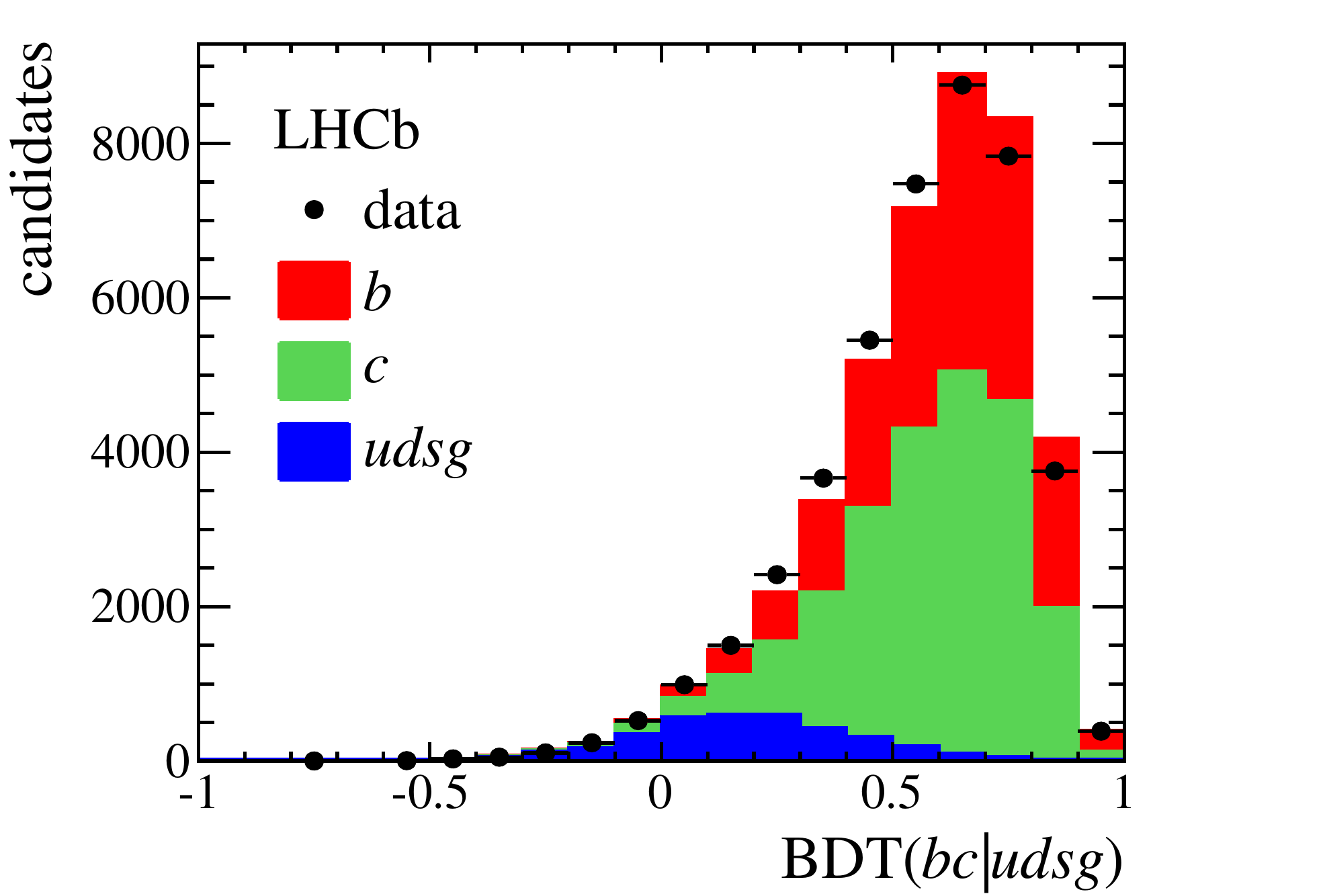}
    \includegraphics[height=4.5cm]{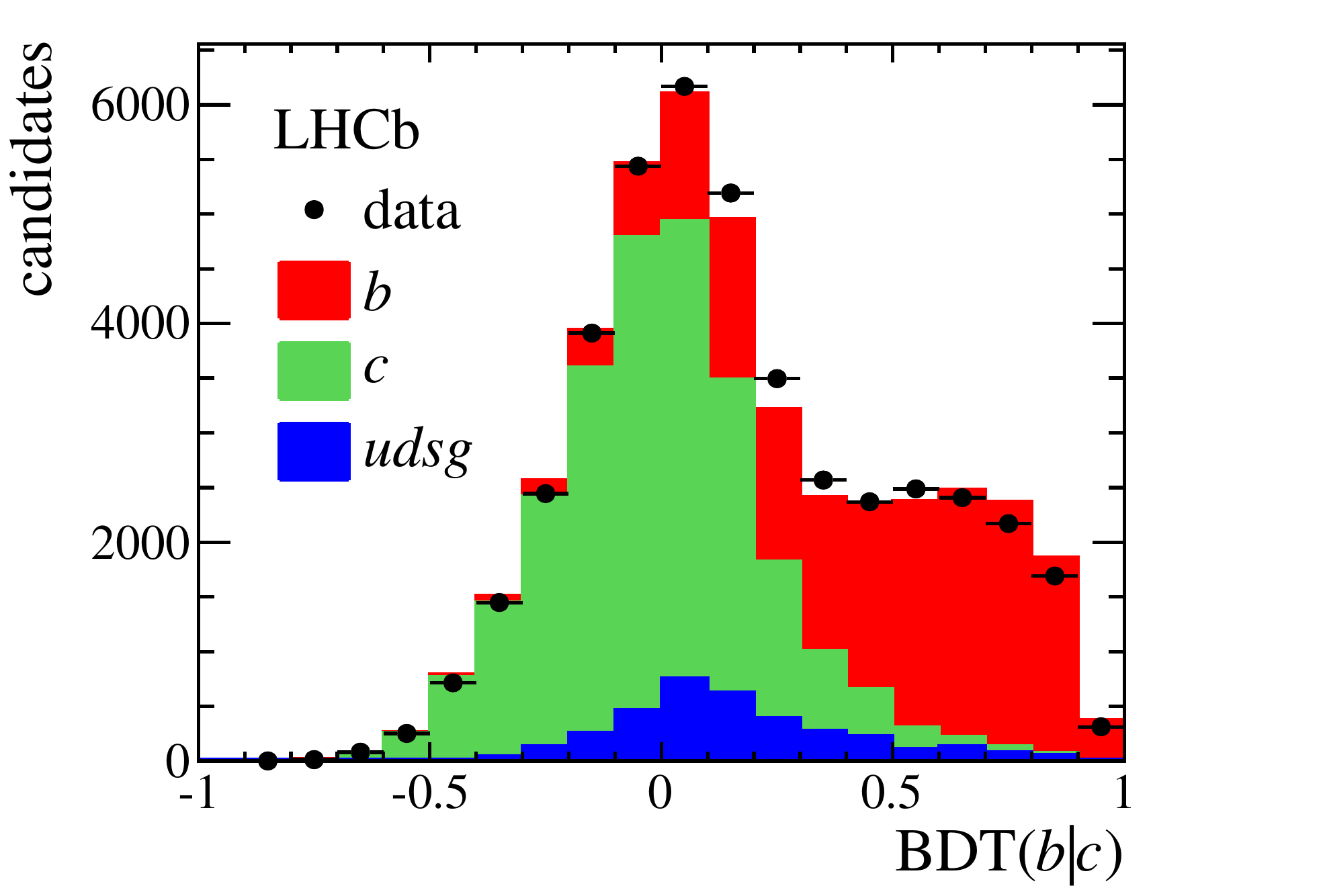}
  \end{center}
  \caption{Projections of the two-dimensional BDT response fit onto
    the (left) \bdtl and (right) \bdth axes for the $c$-hadron
    sample. The stacked fills are (blue) light, (green) $c$, and (red)
    $b$-jet templates from simulation.\label{fig:tag:fits}}
\end{figure}

There is good agreement between the tagging efficiencies obtained from
data and simulation. After SV-tagging a jet, either the BDT
distribution can be fit directly, or further requirements can be
placed on the \bdtl and \bdth responses when high-purity samples are
needed. On the left of Figure~\ref{fig:tag:performance} the efficiency
from data for tagging a $c$ and $b$-jet with an SV is plotted as a
function of jet \pt. By varying the minimum requirement of \bdtl, the
tagging efficiency from simulation as a function of the light-jet
mis-tag rate is given on the right of
Figure~\ref{fig:tag:performance}.

\begin{figure}[t!]
  \begin{center}
    \includegraphics[height=3.9cm]{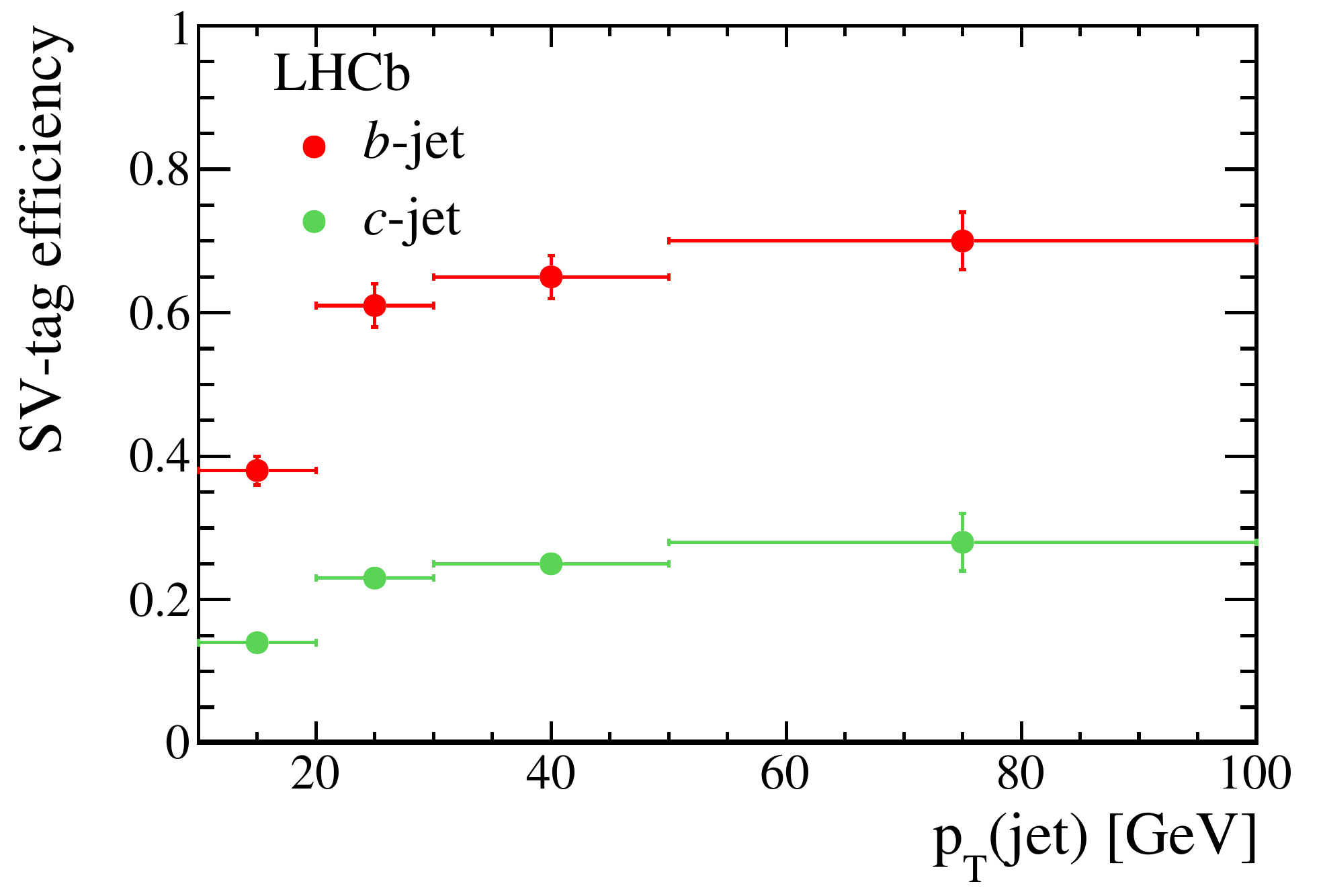}
    \includegraphics[height=4.5cm]{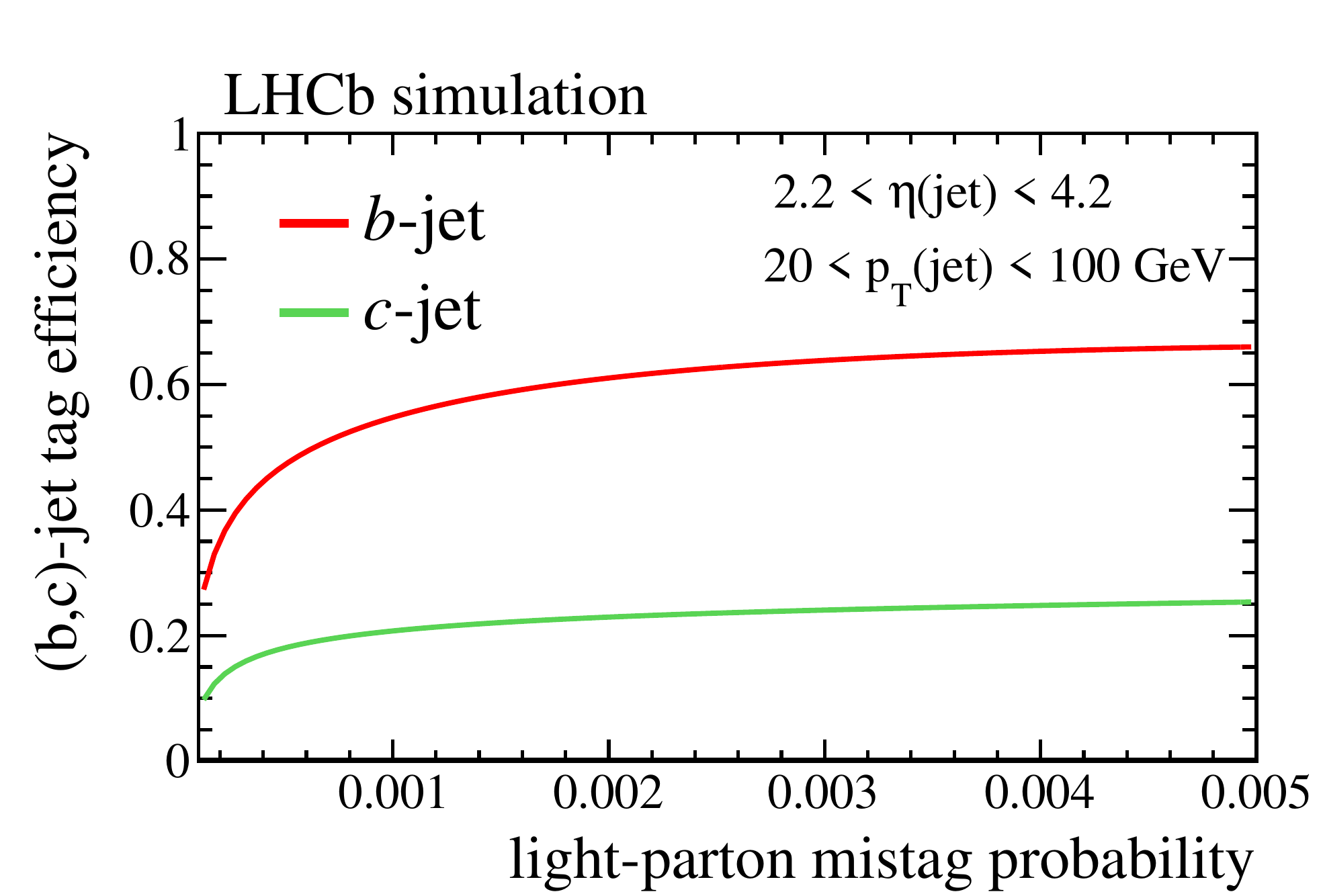}
  \end{center}
  \caption{Tagging efficiency for $c,b$-jets (right) from data as a
    function of jet $\pt$, and (left) from simulation as
    a function of light-jet mis-tag rate.\label{fig:tag:performance}}
\end{figure}

The primary uncertainty on the tagging efficiency is from the
$\log(\ip_{\chi^2})$ fits prior to SV-tagging, and is evaluated by
fixing the light-jet component from the high-\pt muon
sample. Systematic uncertainties from BDT templates, \ip resolution,
muon mis-identification, gluon splitting, and number of $pp$
interactions have also been evaluated. For jets with $\pt > 20\gev$,
the total systematic uncertainty on the tagging efficiency is found to
be $\approx 10\%$ for both $c$ and $b$-jets.


\section[$W+c,b$-jet Ratios]{$\bm{W+c,b}$-jet Ratios}\label{sec:wcb}

\begin{figure}[h!]
  \begin{center}
    \includegraphics[height=4.5cm]{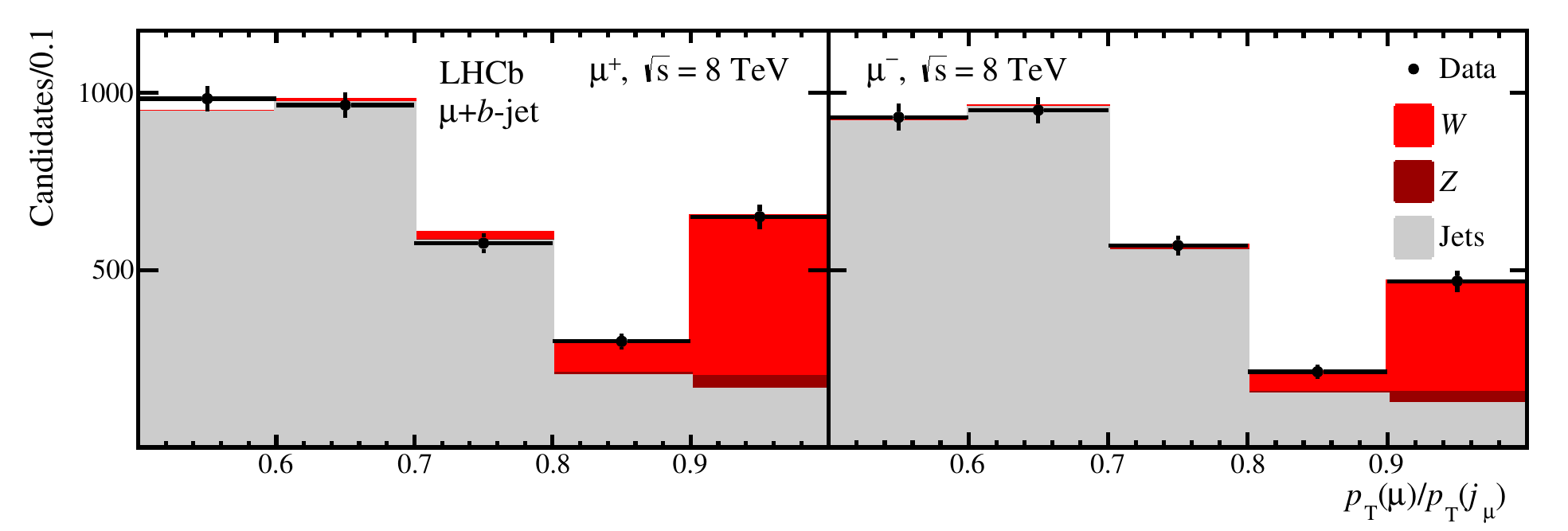}
  \end{center}
  \caption{Isolation distributions and their corresponding template
    fits for the (left) $\mu^++b$-jet and (right) $\mu^-+b$-jet at
    $\sqrt{s} = 8\tev$. The stacked fills are the (gray) di-jet,
    (light red) $W+b$-jet, and (dark red) $Z+b$-jet
    templates.\label{fig:wcb:fits}}
\end{figure}

Measuring $W+c,b$-jet production not only constrains the $s$-quark PDF
of the proton, but also helps determine backgrounds to top production
and understand high-\pt $b$-jet
production. In~\cite{LHCb-PAPER-2015-021} the fiducial definition of a
$W+$jet event requires a muon with $\pt(\mu) > 20\gev$ and $2 <
\eta(\mu) < 4.5$, and a jet with $\pt(j) > 20\gev$ and $2.2 < \eta(j)
< 4.2$. The reduced $\eta$ range of the jet ensures stable
reconstruction and tagging efficiencies. Additionally, the $\Delta R$
between the muon and jet must be greater than $0.5$ and the combined
\pt of the muon and jet, $\pt(\mu+j)$, must be greater than
$20\gev$. Here $\pt(\mu+j) > 20\gev$ is a theoretically well-defined
proxy for the experimental-level selection $\pt(j_\mu+j) > 20\gev$,
where $j_\mu$ is the jet containing the muon. The $\pt(j_\mu+j)$
requirement reduces \pt-balanced di-jet backgrounds, where energy is
not lost to a missing neutrino.

Events are selected by requiring the hardest-\pt muon candidate, and
the hardest-\pt non-muon candidate jet from the same primary vertex,
satisfy all fiducial requirements with the substitution $\pt(j_\mu+j)$
for $\pt(\mu+j)$. Events are binned as a function of isolation,
$\pt(\mu)/\pt(j_\mu)$. The $\mu+c,b$ content of each bin is determined
by requiring only events with an SV-tagged jet and performing the BDT
fit of Figure~\ref{fig:tag:fits}. The isolation distribution of the
full sample is fit to determine the $W+$jet yield, while the
$\mu+c,b$-jet distributions are fit to determine the $W+c,b$-jet
yields. The fits are split by muon charge and performed separately for
the $7\tev$ and $8\tev$ datasets. In Figure \ref{fig:wcb:fits} the
$\mu+c,b$-jet isolation distribution fits at $\sqrt{s} = 8\tev$ are
provided, where good agreement can be seen between the data and fit.

The measured $W+c,b$-jet asymmetries, $W+c,b$-jet to $W+$jet ratios,
and $W+$jet to $Z+$jet ratios are,
\begin{center}
  \def\svgwidth{\columnwidth}%
  \executeiffilenewer{wcb.svg}{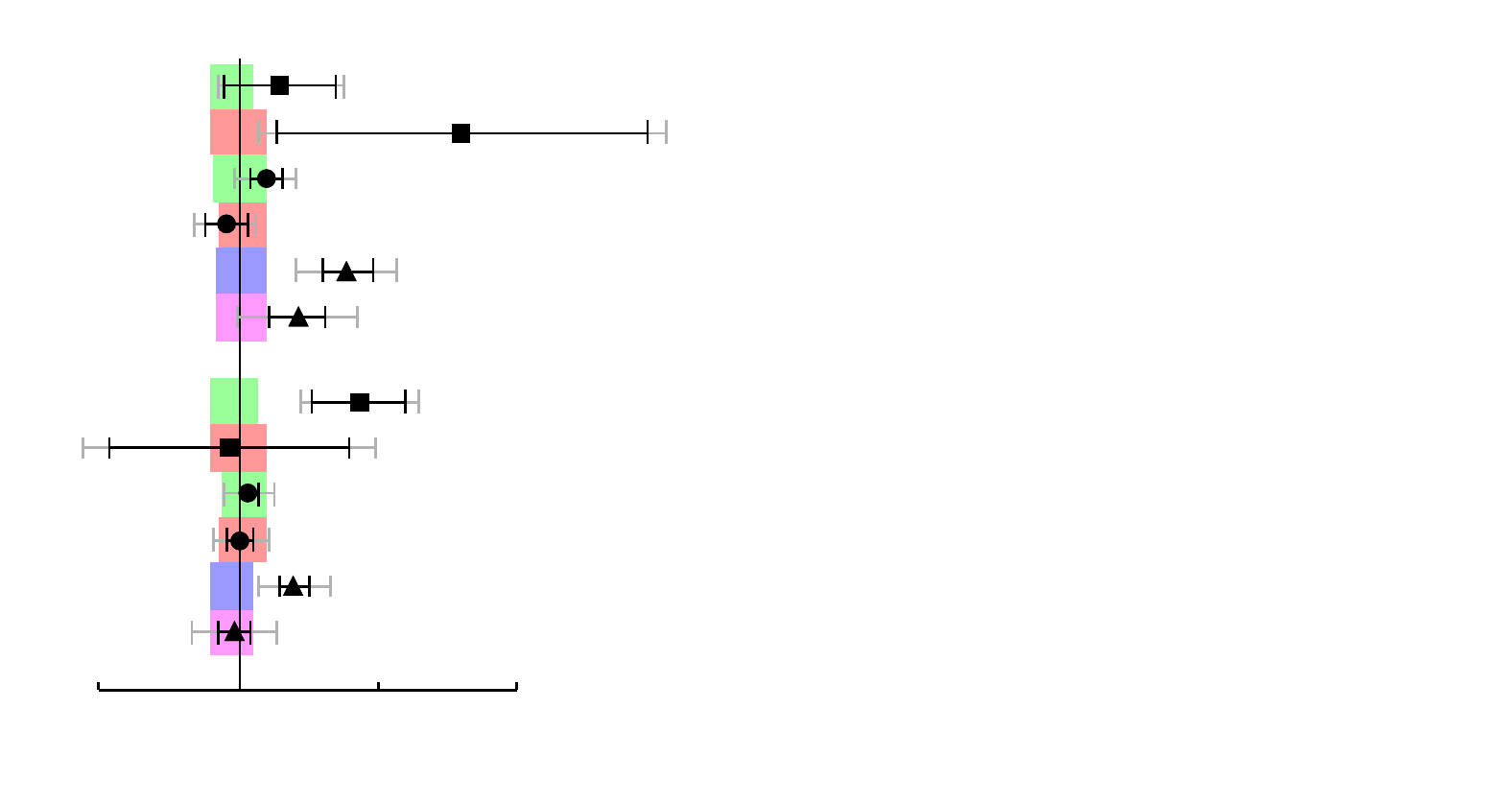}%
  {inkscape -z -D --file=wcb.svg %
    --export-pdf=wcb.pdf --export-latex}%
\begingroup%
  \makeatletter%
  \providecommand\color[2][]{%
    \errmessage{(Inkscape) Color is used for the text in Inkscape, but the package 'color.sty' is not loaded}%
    \renewcommand\color[2][]{}%
  }%
  \providecommand\transparent[1]{%
    \errmessage{(Inkscape) Transparency is used (non-zero) for the text in Inkscape, but the package 'transparent.sty' is not loaded}%
    \renewcommand\transparent[1]{}%
  }%
  \providecommand\rotatebox[2]{#2}%
  \ifx\svgwidth\undefined%
    \setlength{\unitlength}{453.6bp}%
    \ifx\svgscale\undefined%
      \relax%
    \else%
      \setlength{\unitlength}{\unitlength * \real{\svgscale}}%
    \fi%
  \else%
    \setlength{\unitlength}{\svgwidth}%
  \fi%
  \global\let\svgwidth\undefined%
  \global\let\svgscale\undefined%
  \makeatother%
  \begin{picture}(1,0.5308642)%
    \put(0,0){\includegraphics[width=\unitlength]{wcb.pdf}}%
    \put(0.68077601,0.46737213){\color[rgb]{0,0,0}\makebox(0,0)[rb]{\smash{$-0.09\pm0.08\pm0.04$}}}%
    \put(0.70017637,0.46737213){\color[rgb]{0,0,0}\makebox(0,0)[lb]{\smash{$\mathcal{A}(Wc)$}}}%
    \put(0.68077601,0.4356261){\color[rgb]{0,0,0}\makebox(0,0)[rb]{\smash{$0.51\pm0.20\pm0.09$}}}%
    \put(0.70017637,0.4356261){\color[rgb]{0,0,0}\makebox(0,0)[lb]{\smash{$\mathcal{A}(Wb)$}}}%
    \put(0.68077601,0.40564374){\color[rgb]{0,0,0}\makebox(0,0)[rb]{\smash{$5.80\pm0.44\pm0.75$}}}%
    \put(0.70017637,0.40564374){\color[rgb]{0,0,0}\makebox(0,0)[lb]{\smash{$\sigma(Wc)/\sigma(Wj)\times10^2$}}}%
    \put(0.68077601,0.37566138){\color[rgb]{0,0,0}\makebox(0,0)[rb]{\smash{$0.66\pm0.13\pm0.13$}}}%
    \put(0.70017637,0.37566138){\color[rgb]{0,0,0}\makebox(0,0)[lb]{\smash{$\sigma(Wb)/\sigma(Wj)\times10^2$}}}%
    \put(0.68077601,0.34391534){\color[rgb]{0,0,0}\makebox(0,0)[rb]{\smash{$6.61\pm0.19\pm0.33$}}}%
    \put(0.70017637,0.34391534){\color[rgb]{0,0,0}\makebox(0,0)[lb]{\smash{$\sigma(W^-j)/\sigma(Zj)$}}}%
    \put(0.68077601,0.31393298){\color[rgb]{0,0,0}\makebox(0,0)[rb]{\smash{$10.49\pm0.28\pm0.53$}}}%
    \put(0.70017637,0.31393298){\color[rgb]{0,0,0}\makebox(0,0)[lb]{\smash{$\sigma(W^+j)/\sigma(Zj)$}}}%
    \put(0.68077601,0.25749559){\color[rgb]{0,0,0}\makebox(0,0)[rb]{\smash{$-0.01\pm0.05\pm0.04$}}}%
    \put(0.70017637,0.25749559){\color[rgb]{0,0,0}\makebox(0,0)[lb]{\smash{$\mathcal{A}(Wc)$}}}%
    \put(0.68077601,0.22751323){\color[rgb]{0,0,0}\makebox(0,0)[rb]{\smash{$0.27\pm0.13\pm0.09$}}}%
    \put(0.70017637,0.22751323){\color[rgb]{0,0,0}\makebox(0,0)[lb]{\smash{$\mathcal{A}(Wb)$}}}%
    \put(0.68077601,0.19753086){\color[rgb]{0,0,0}\makebox(0,0)[rb]{\smash{$5.62\pm0.28\pm0.73$}}}%
    \put(0.70017637,0.19753086){\color[rgb]{0,0,0}\makebox(0,0)[lb]{\smash{$\sigma(Wc)/\sigma(Wj)\times10^2$}}}%
    \put(0.68077601,0.16578483){\color[rgb]{0,0,0}\makebox(0,0)[rb]{\smash{$0.78\pm0.08\pm0.16$}}}%
    \put(0.70017637,0.16578483){\color[rgb]{0,0,0}\makebox(0,0)[lb]{\smash{$\sigma(Wb)/\sigma(Wj)\times10^2$}}}%
    \put(0.68077601,0.13580247){\color[rgb]{0,0,0}\makebox(0,0)[rb]{\smash{$6.02\pm0.13\pm0.30$}}}%
    \put(0.70017637,0.13580247){\color[rgb]{0,0,0}\makebox(0,0)[lb]{\smash{$\sigma(W^-j)/\sigma(Zj)$}}}%
    \put(0.68077601,0.10582011){\color[rgb]{0,0,0}\makebox(0,0)[rb]{\smash{$9.44\pm0.19\pm0.47$}}}%
    \put(0.70017637,0.10582011){\color[rgb]{0,0,0}\makebox(0,0)[lb]{\smash{$\sigma(W^+j)/\sigma(Zj)$}}}%
    \put(0.29982363,0.38800705){\color[rgb]{0,0,0}\makebox(0,0)[lb]{\smash{7\tev}}}%
    \put(0.29982363,0.17813051){\color[rgb]{0,0,0}\makebox(0,0)[lb]{\smash{8\tev}}}%
    \put(0.20458554,0.01763668){\color[rgb]{0,0,0}\makebox(0,0)[b]{\smash{(exp - thr)/$\max(\delta_\mathrm{thr})$}}}%
    \put(0.06525573,0.04409171){\color[rgb]{0,0,0}\makebox(0,0)[b]{\smash{-5}}}%
    \put(0.15873016,0.04409171){\color[rgb]{0,0,0}\makebox(0,0)[b]{\smash{0}}}%
    \put(0.25044092,0.04409171){\color[rgb]{0,0,0}\makebox(0,0)[b]{\smash{5}}}%
    \put(0.34215168,0.04409171){\color[rgb]{0,0,0}\makebox(0,0)[b]{\smash{10}}}%
  \end{picture}%
\endgroup%

\end{center}
where the first uncertainty is statistical and the second is
systematic. The asymmetry $\mathcal{A}(Wq)$ is defined as
$(\sigma(W^+q) - \sigma(W^-q))/(\sigma(W^+q) + \sigma(W^-q))$.

All measurements are unitless. Each observable is graphically compared
to its SM prediction using the difference between experiment and
theory over the maximum theory uncertainty; the points are this
quantity, the gray and black bars are the total and statistical
experimental uncertainties, and the asymmetric colored bands are the
theory uncertainties. The SM predictions are calculated with the
four-flavor scheme at NLO using MCFM~\cite{Campbell:2000bg} and the
CT10 PDF set~\cite{Lai:2010vv}, where the total uncertainty is the
combined PDF, $\alpha_s$, and scale uncertainty.

Because these measurements are ratios, most reconstruction
efficiencies cancel. However, the $c,b$-tagging efficiencies, taken
from Section~\ref{sec:tag}, enter the $\sigma(Wc,b)/\sigma(Wj)$
ratios. For the $\sigma(Wc)/\sigma(Wj)$ ratio, the $c$-tagging
efficiency is the primary systematic uncertainty, while the
subtraction of top backgrounds from a sideband is the primary
uncertainty for the $\sigma(Wb)/\sigma(Wj)$ measurement. Backgrounds
from $\tau$ decays are subtracted from the $W+c$-jet measurements but
are negligible. The primary uncertainty on both the asymmetries and
the $Wj/Zj$ ratios is from the isolation fits.

\section{Top Cross-Section}\label{sec:top}

A tightened fiducial region of $\pt(\mu) > 25\gev$ and $50 < \pt(j) <
100$ is applied to the analysis of Section~\ref{sec:wcb} in
\cite{LHCb-PAPER-2015-022} to obtain a top-quark enriched data sample;
the top quarks are from both single top ($\approx 25\%$) and top-pair
production ($\approx 75\%$). The additional muon requirement reduces
the di-jet background, while the jet requirement suppresses direct
$W+b$-jet background. The jet is required to be SV-tagged and the
$\mu+b$-jet yield is determined from the isolation distribution via
the methods of Section~\ref{sec:wcb}.

Despite the increased jet \pt requirement, a sizable background from
direct $W+b$-jet production, \textit{i.e.} not from top, remains in
the $W+b$-jet yield. This background is constrained by determining the
$W+$jet yield from data without an SV-tag, applying the $b$-tag
efficiency, and correcting with the ratio $\sigma(Wb)/\sigma(Wj)$ from
theory. Here, the theoretical uncertainty on $\sigma(Wb)/\sigma(Wj)$
is considerably smaller than for $\sigma(Wb)$ alone. This method is
cross-checked against the $W+c$-jet yield, where no top production is
present, and is found to describe the data well.

\begin{figure}
  \begin{center}
    \includegraphics[height=4.5cm]{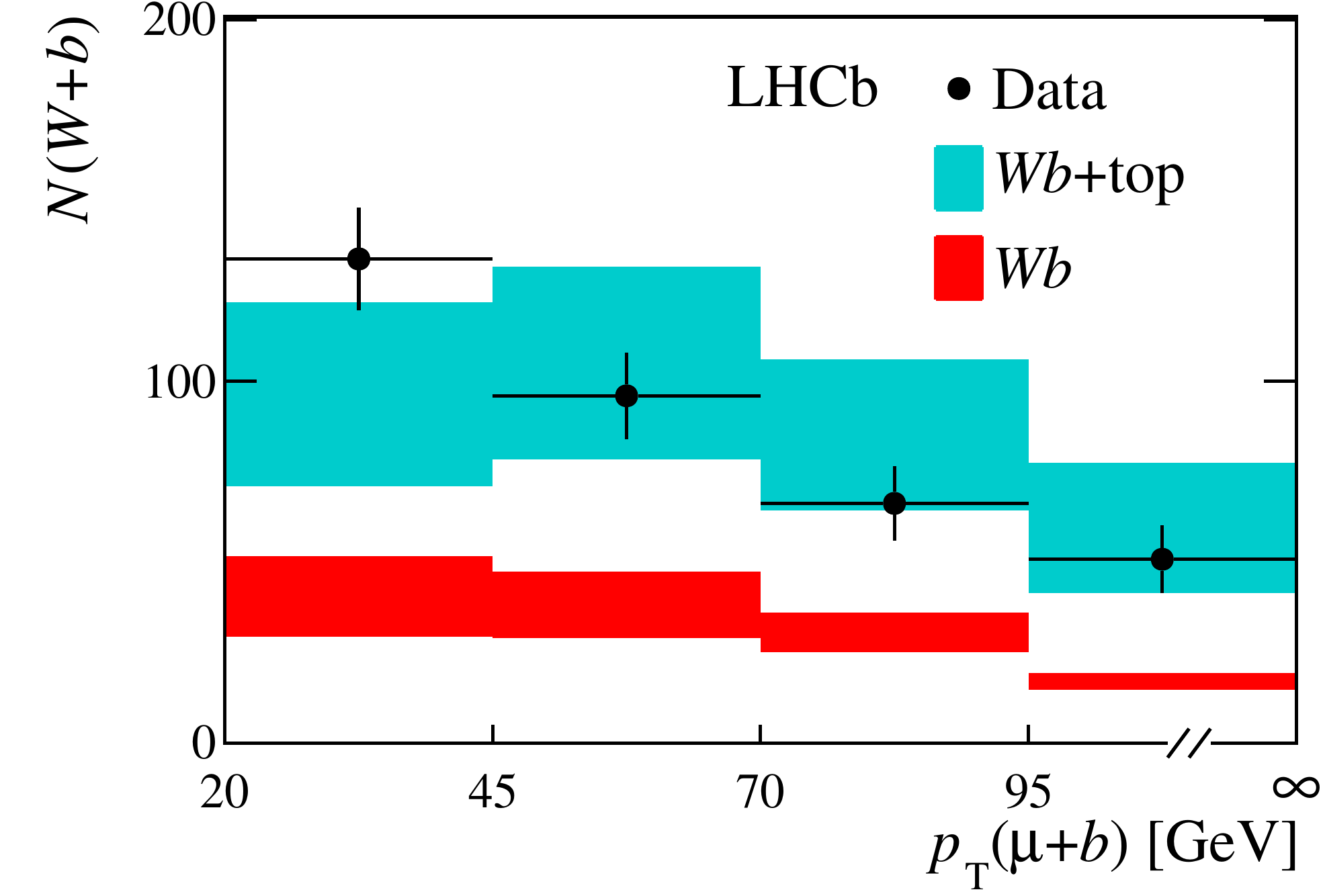}
    \includegraphics[height=4.5cm]{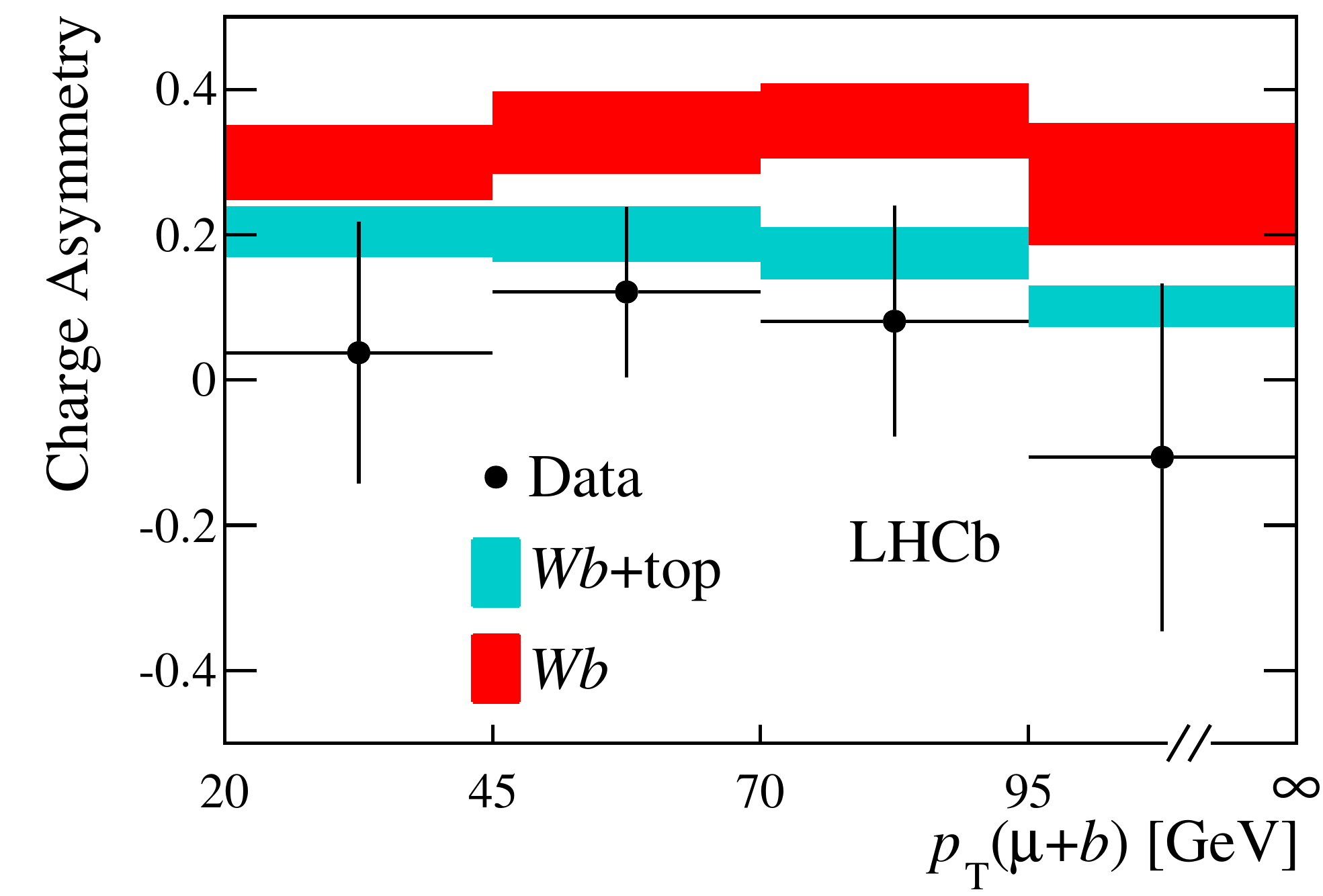}
  \end{center}
  \caption{Combined $7$ and $8\tev$  distributions of (left) the
    number of events as a function of $\pt(\mu+b)$ and (right) the
    asymmetry, also as a function of $\pt(\mu+b)$. The points are
    data, while the fills are the SM predictions for (red) $W+b$-jet
    production without top and (cyan) $W+b$-jet production with
    top.\label{fig:top:fits}}
\end{figure}

In the left plot of Figure~\ref{fig:top:fits} the $W+b$-jet yield from
the combined $7$ and $8\tev$ data is plotted as a function of the muon
and $b$-jet \pt. The red band, with uncertainty, is the constrained
$W+b$-jet prediction without top, while the cyan band, also with
uncertainty, includes the SM top prediction. The yield, particularly
at high $\pt(\mu+b)$ cannot be described by direct $W+b$-jet
production alone. Similarly, the asymmetry as a function of
$\pt(\mu+b)$ is plotted on the left of
Figure~\ref{fig:top:fits}. Here, the direct $W+b$-jet asymmetry is
near $\approx 1/3$ due to valence quark content, while the top-pair
asymmetry is $\approx 0$. Again, the direct $W+b$-jet hypothesis
without top production does not describe the data well.

A binned profile likelihood fit of these two distributions is
performed with the top contribution allowed to vary freely. Systematic
uncertainties, both theoretical and experimental, are introduced as
Gaussian nuisance parameters, and the SM hypothesis with and without
top is compared. A $5.4\sigma$ significance is observed, indicating
the presence of top production in the forward region. The top yield is
then determined by subtracting the direct $W+b$-jet contribution
constrained from data.

Correcting for reconstruction efficiencies, the $7$ and $8\tev$
measured top cross-sections are,
\begin{center}
  \def\svgwidth{0.75\columnwidth}%
  \executeiffilenewer{top.svg}{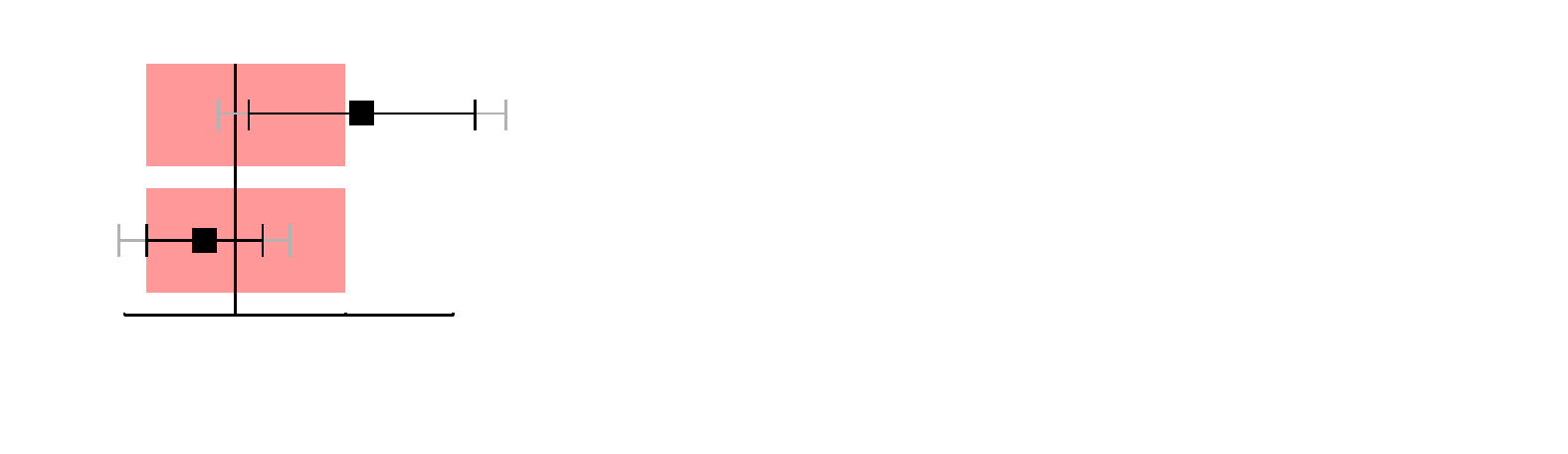}%
  {inkscape -z -D --file=top.svg %
    --export-pdf=top.pdf --export-latex}%
\begingroup%
  \makeatletter%
  \providecommand\color[2][]{%
    \errmessage{(Inkscape) Color is used for the text in Inkscape, but the package 'color.sty' is not loaded}%
    \renewcommand\color[2][]{}%
  }%
  \providecommand\transparent[1]{%
    \errmessage{(Inkscape) Transparency is used (non-zero) for the text in Inkscape, but the package 'transparent.sty' is not loaded}%
    \renewcommand\transparent[1]{}%
  }%
  \providecommand\rotatebox[2]{#2}%
  \ifx\svgwidth\undefined%
    \setlength{\unitlength}{453.6bp}%
    \ifx\svgscale\undefined%
      \relax%
    \else%
      \setlength{\unitlength}{\unitlength * \real{\svgscale}}%
    \fi%
  \else%
    \setlength{\unitlength}{\svgwidth}%
  \fi%
  \global\let\svgwidth\undefined%
  \global\let\svgscale\undefined%
  \makeatother%
  \begin{picture}(1,0.28924162)%
    \put(0,0){\includegraphics[width=\unitlength]{top.pdf}}%
    \put(0.85008818,0.20282187){\color[rgb]{0,0,0}\makebox(0,0)[rb]{\smash{$239\pm53\pm41$ [fb]}}}%
    \put(0.33862434,0.20282187){\color[rgb]{0,0,0}\makebox(0,0)[lb]{\smash{7\tev}}}%
    \put(0.85008818,0.12345679){\color[rgb]{0,0,0}\makebox(0,0)[rb]{\smash{$289\pm43\pm49$ [fb]}}}%
    \put(0.33862434,0.12345679){\color[rgb]{0,0,0}\makebox(0,0)[lb]{\smash{8\tev}}}%
    \put(0.18518519,0.00881834){\color[rgb]{0,0,0}\makebox(0,0)[b]{\smash{(exp - thr)/$\max(\delta_\mathrm{thr})$}}}%
    \put(0.07936508,0.05114638){\color[rgb]{0,0,0}\makebox(0,0)[b]{\smash{-1}}}%
    \put(0.14991182,0.05114638){\color[rgb]{0,0,0}\makebox(0,0)[b]{\smash{0}}}%
    \put(0.22045855,0.05114638){\color[rgb]{0,0,0}\makebox(0,0)[b]{\smash{1}}}%
    \put(0.28924162,0.05114638){\color[rgb]{0,0,0}\makebox(0,0)[b]{\smash{2}}}%
  \end{picture}%
\endgroup%

\end{center}
where the first uncertainty is statistical and the second is the
combined experimental and theoretical systematic uncertainties. Just
as for the $W+c,b$-observables, each top cross-section is also
graphically compared to its corresponding SM prediction calculated at
NLO using MCFM with the four-flavor scheme. The primary systematic
uncertainty is from the $b$-tagging efficiency, but the systematic
uncertainties between the $7$ and $8\tev$ measurements are nearly
completely correlated.

\section{Conclusion}

Inclusive $c$ and $b$-jet tagging has been developed and validated
using run 1 data from LHCb. This tagging in turn has been used to
measure $W+c,b$-jet ratios and asymmetries as well as forward top
production cross-sections. With significantly increased statistics
during run 2 of the LHC, updates of these measurements will have
significant physics impact, including constraining both $s$-quark and
gluon PDFs, probing intrinsic $b$-content, and even possibly measuring
the non-zero top-pair asymmetry. Further studies are underway to
further improve tagging efficiencies as well as determine physics
measurements that can utilize inclusive $c$-tagging.

\setboolean{inbibliography}{true}
\bibliographystyle{lhcb}
\bibliography{proceedings}

\end{document}